\documentclass{mem}
\usepackage{natbib}\usepackage{txfonts}\usepackage{balance}
\usepackage{graphicx}
\usepackage[a4paper]{hyperref}
\idline{1}{1}
\begin{document}
\def\teff{$T\rm_{eff }$}
\def\kms{$\mathrm {km s}^{-1}$}
\newcommand{\Hi}{\textup{H\,{\mdseries\textsc{i}}}}
\newcommand{\HI}{\textup{H\,{\mdseries\textsc{i}}}~}
\newcommand{\HII}{\textup{H\,{\mdseries\textsc{ii}}}~}
\newcommand{\Kkms}{{K~km~s$^{-1}$}}
\newcommand{\Htwo}{{H$_{2}$}\ }
\newcommand{\coa}{{$^{12}$CO(1-0)}}
\newcommand{\cob}{{$^{12}$CO(2-1)}}
\newcommand{\Msun}{{M$_{\odot}$}}
\newcommand{\Lsun}{{L$_{\odot}$}}
\newcommand{\MHtwo}{{M$_{\rm H_2}$}}
\newcommand{\MHI}{{M$_{\rm H_I}$}}
\newcommand{\column}{{$10^{20}$~cm$^{-2}$~(K~km~s$^{-1}$)$^{-1}$}}
\newcommand{\vrel}{{v$_{\rm rel}$}}
\newcommand{\Rc}{{R$_{\rm c}$}}
\newcommand{\dv}{{$\Delta \rm v$}}
\newcommand{\Mvir}{{M$_{ \rm vir}$}}
\def\kms{{km~s$^{-1}$}}
\def\fmag{\hbox{$.\!\!^{\rm m}$}}
\def\degr{\hbox{$^\circ$}}
\def\arcmin{\hbox{$^\prime$}}
\def\arcsec{\hbox{$^{\prime\prime}$}}
\def\fdegr{\hbox{$.\!\!^\circ$}}
\def\farcmin{\hbox{$.\mkern-4mu^\prime$}}
\def\farcsec{\hbox{$.\!\!^{\prime\prime}$}}

\title{
Circumnuclear HI disks in radio galaxies
}

   \subtitle{The case of Cen~A and B2~0258+35}

\author{
C. \,Struve\inst{1,2},
R. \,Morganti\inst{1,2} 
\and T.A. \,Oosterloo\inst{1,2}
          }


\institute{
Netherlands Foundation for Research in Astronomy,
Postbus 2, 7990~AA, Dwingeloo, The Netherlands
\and
Kapteyn Institute, University of Groningen,
Landleven 12, 9747~AD, Groningen, The Netherlands\\
\email{struve@astron.nl}
}

\authorrunning{Struve et al. }

\titlerunning{Circumnuclear \HI disks in radio galaxies}

\abstract{
New \HI observations of the nearby radio-loud galaxies Centaurus~A and B2~0258+35 show broad absorption ($\Delta v_{\rm{absorp}}\sim 400$~\kms ) against the unresolved nuclei. Both sources belong to the cases where blue- and redshifted absorption is observed at the same time. In previous Cen~A observations only a relative narrow range of redshifted absorption was detected.
We show that the data suggest in both cases the existence of a circumnuclear disk. For Cen~A the nuclear absorption might be the atomic counterpart of the molecular circumnuclear disk that is seen in CO and H$_2$. Higher resolution observations are now needed to locate the absorption and to further investigate the structure and kinematics of the central region of the AGN and the way the AGN are fueled.
\keywords{galaxies: active -- galaxies: ISM -- galaxies: individual: Centaurus~A and B2~0258+35 -- galaxies: kinematics and dynamics }
}
\maketitle{}
\section{Introduction}
The central region of AGN and the way AGN are fueled are important topics in current extra-galactic astronomy. Mergers have always been considered important for the triggering of AGN by providing the mechanism that could bring gas to the central regions and fuel the black hole \citep[e.g.][]{hibbard96,mihos96,barnes02}. While this seems to be likely in powerful radio galaxies, recent observations show that other AGN - e.g. low-power radio galaxies - exist where there is no evidence for accretion through mergers. The activity in these galaxies appears to be associated with the slow accretion of gas from the ISM/IGM \citep{best05,balmaverde08}.\\
A powerful tool to learn more about the structure of the central regions of AGN and the fueling mechanism is the study of the distribution and kinematics of the gas. In order to perform detailed studies of single objects we have selected two \Hi -rich radio-loud galaxies, the merger remnant Centaurus~A and B2~0258+35 (NGC~1167). The later is unlikely to be a recent merger but has a young radio source suggesting that ``cold accretion'' from the IGM plays the dominant role by providing the fuel for the AGN activity (see Sect.~3).\\
Neutral hydrogen is an ideal tracer of accretion, interaction and merging events. \HI in absorption has been often detected in radio sources \citep[see e.g.][]{conway97,vermeulen03,morganti05}. These absorption structures have shown a variety of characteristics. \HI absorption profiles only redshifted (relative to the systemic velocity), were initially found in a number of radio galaxies and interpreted as gas clouds that are falling into the nucleus, possibly indicating accretion of gas which deliver the fuel for the AGN \citep[e.g.][]{vangorkom89}. However, more sensitive and broader band observations have shown that the picture is more complicated and that blueshifted absorption occurs even more often than redshifted absorption \citep[e.g.][]{vermeulen03,morganti05}.\\
In other cases, the \HI absorption is centered on the systemic velocity of the galaxy and these cases have been often interpreted as \HI associated with a circumnuclear disk (or torus), see e.g. \citet{conway97}, \citet{peck01}. Support for the picture of a circumnuclear disk comes from theoretical work that indeed predict the existence of such circumnuclear \HI disks around (active) black holes \citep[see e.g.][]{maloney96,loeb08}. Thus, the picture is more complicated and the questions concerning the central structure and fueling remains.
\begin{figure*}[t!]
\includegraphics[height=4.6cm]{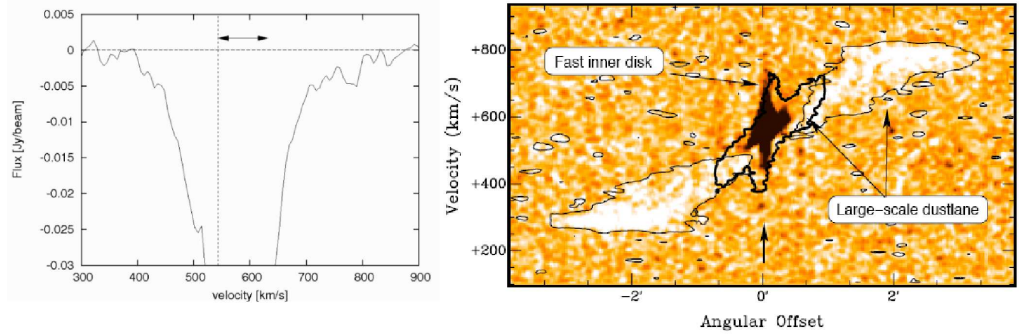}
\caption{\footnotesize
Centaurus~A: Left panel: \HI absorption against the unresolved nucleus as derived from the ATCA observations \citep{morganti08}. The total absorption width ranges from 400 to 800~\kms . The arrow indicates the previously known (redshifted) absorption. The vertical line gives the systemic velocity. Right panel: Position-velocity plot of the \HI (grey-scale and thin contours) and superimposed the CO emission (thick contours; from \citet{liszt01}, taken along position angle 139\degr . Note that the CO observations do not extend beyond a radius of about 1 arcmin. Figure taken from \citet{morganti08}.}
\label{cena}
\end{figure*}
\section{Centaurus~A}
The prime galaxy to investigate these questions is Centaurus~A, the closest radio-loud galaxy at a distance of only 3.5~Mpc. Observations have shown redshifted \HI absorption against the unresolved nucleus \citep[e.g.][]{vdH83,sarma02}. This has been interpreted as gas clouds close to the nucleus which are falling in, feeding the black hole.\\
However, our new ATCA observations (Morganti et al. 2008; Struve et al. in prep.) revealed for the first time blueshifted absorption (w.r.t. the systemic velocity). Moreover, the data show that the nuclear redshifted absorption is significantly broader than reported before, see Fig.~1. The total absorption width is 400~\kms .\\
The large-scale gas and dust disk has a complicated warped structure which is close to edge-on. Thanks to a careful modeling of the warped disk of \HI seen in emission (and partly in absorption) we have shown that the nuclear absorption must arise mainly in the vicinity of the nucleus (Struve et al. in prep.). This raises the question of what the structure in the vicinity of the black hole is and which mechanism fuels the black hole.\\
The fact that red- and blueshifted absorption is observed at the same time suggests that the \HI absorption is not, as was previously claimed, evidence of gas infall into the AGN, but instead is due to a cold, circumnuclear rotating disk, similar to that seen in CO \citep[Fig.~\ref{cena},][]{liszt01} and at other wavelengths \citep[e.g.][]{neumayer08}. However, the absorption is limited by the background structure. VLBI observations, which have much higher angular resolution than the ATCA data (beam size $\sim 6$\arcsec ), revealed significant structure on milli-arcsecond scale: an unresolved nucleus, a jet and a counterjet \citep[e.g.][]{jones96,tingay01}. The nucleus is only visible at high frequencies, while at low frequencies \citep[2.3~GHz][]{jones96} the VLBI nucleus is obscured, suggesting that the spectrum is inverted and the nucleus is also invisible at 1.4~GHz. Thus, the absorption observed with the ATCA must arise from an extended structure.\\
The nuclear disk detected in molecular and ionized-gas lines \citep[see ref. in][]{morganti08} has a major axis that is roughly perpendicular to the jet axis. Therefore - if the \HI is located in a rotating disk and if the absorption arises against the jet/counter-jet - only velocities close to the systemic are expected, given the small opening angle of the jet. Thus, it would be hard to explain the large velocity width. The absorption is likely to be occurring elsewhere. Interestingly, extra flux ($>50$~mJy) is clearly present on the short baselines in the 1.4~GHz VLBI experiments \citep{tingay01}, indicating the presence of continuum structures that are too large to be properly imaged. Therefore, the absorption could (at least partly) also arise in front of this diffuse component, explaining the full absorption width observed in the case of a circumnuclear \HI disk.\\
\begin{figure*}[t!]
\includegraphics[height=4.7cm]{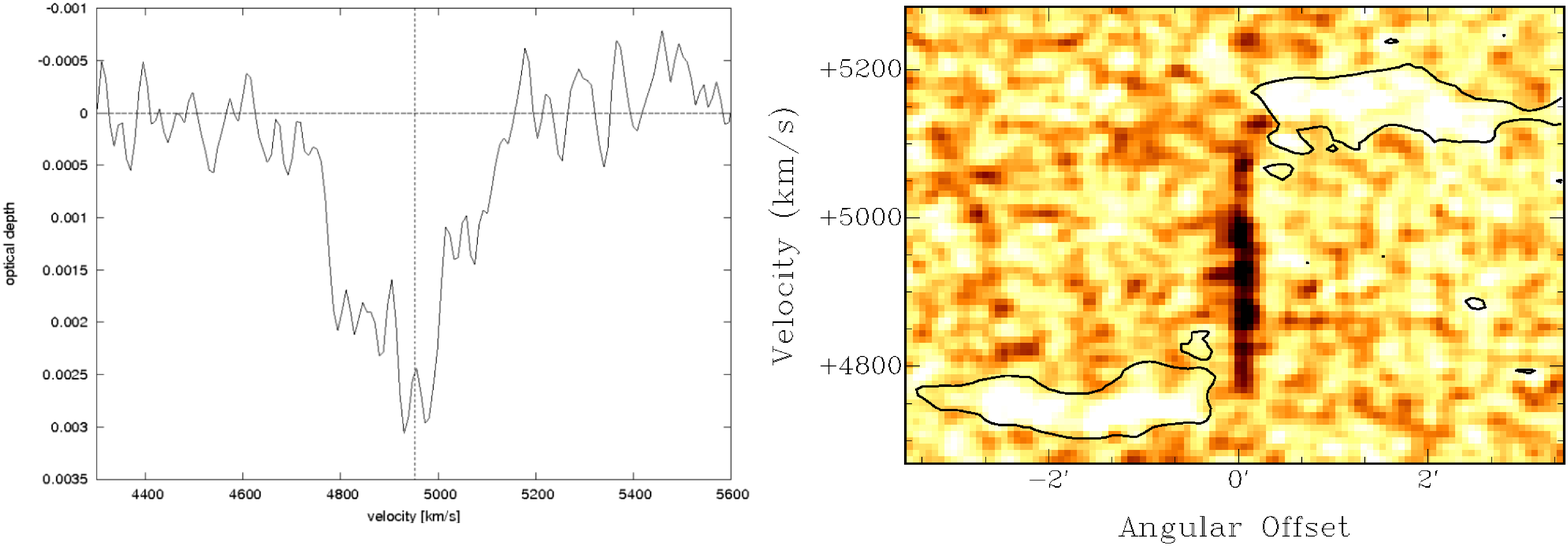}
\caption{\footnotesize
B2~0259+35: Left panel: \HI absorption against the unresolved nucleus as derived from the new WSRT observations (Struve et al. in prep.). The total absorption width is larger than 400~\kms . The vertical line indicates the systemic velocity. Right panel: Position-velocity plot of the \HI (grey-scale and contours) taken along the major axis ($\rm{PA}=75$\degr ).}
\label{0258}
\end{figure*}
\section{B2~0258+35}
B2~0258+35 is a young radio source \citep[$\sim 10^6$~yr,][]{giroletti05} with a two-lobes structure extended about 1~kpc. The stellar population contains old plus intermediate age ($>4$~Gyr) stars, but does not have a young component \citep{emonts06}, suggesting that the host galaxy NGC~1167 is not a recent merger. Therefore, the radio source is unlikely to be triggered by a merging event.  A large ($\sim 160$ kpc), regularly rotating \HI disk (containing more than $10^{10}$ M$_\odot$ of \HI) has been observed with the WSRT \citep{emonts06}. New WSRT observations show the presence of halo gas which is being accreted from the IGM (Struve et al. in prep.).  The large, regular rotating \HI disk (which is not warped) further confirms that NGC~1167 is an old merger but the presence of halo gas suggests that ``cold accretion'' from IGM may be occurring, and could perhaps be related to the triggering of the AGN.\\
Against the radio source (unresolved in the WSRT observations) fairly symmetric blue- and redshifted absorption is found, $\Delta v_{\rm{absorp}} > 400$~\kms ~(Fig.~\ref{0258}, Struve et al. in prep.). Unlike for Cen~A, the continuum structure of B2~0258+35 is extended around the nucleus on the arcsecond scale in all directions and not only along the jet axis \citep{giroletti05}. Hence, the data suggests that the absorption is caused by a cold, circumnuclear disk.\\
\section{Conclusions and Outlook}
Using new, broad band observations of Cen~A and B2~0258+35 we have shown the presence of blue- and redshifted \HI absorption against the unresolved nuclei. Both galaxies belong to the group of radio-loud galaxies where blue- and redshifted absorption is observed at the same time. Previously, for Cen~A only a narrow range of redshifted absorption was detected.\\
We have shown \citep{morganti08} that the \HI absorption in Cen~A is not evidence of gas infall into the AGN, but instead is due to a cold, circumnuclear rotating disk. The absorption profile of the WSRT data suggests the same interpretation for B2~0258+35. Both observations are in agreement with theoretical work which predicts the existence of such circumnuclear disks. However, these results leave the fueling question open.\\
Sensitive, higher resolution observations (VLBI for Cen~A and VLA A-array for B2~0258+35) are now needed (and proposed) to detect and locate the absorption in order to further explore the characteristics of the nuclear \HI and to test the hypotheses of circumnuclear disks.\\
Both sources are unique objects. Cen~A is the closest AGN and therefore the linear scale is much better than for any other radio-loud source. Hence, the inner structure and fueling mechanism can be studied in great detail. Because the background structure of B2~0258+35 is extended around all directions of the black hole, higher spatial resolution data will allow to investigate the complete 2-D kinematics in the vicinity of the AGN. For both sources it will be interesting to investigate the presence of velocity gradients (which might be evidence for circumnuclear disks) as well as radial motions (infall/outflow?).
\begin{acknowledgements}
This research was supported by the EU Framework 6 Marie Curie Early Stage Training programme under contract number MEST-CT-2005-19669 "ESTRELA". The Australia Telescope is funded by the Commonwealth of Australia for operation as a National Facility managed by CSIRO. The WSRT is operated by the Netherlands Foundation for Research in Astronomy with support from the Netherlands Foundation for Scientific Research.
\end{acknowledgements}

\bibliographystyle{aa}

\end{document}